\documentclass{iopconfser}

\usepackage{amssymb,amsmath,amsthm,amsbsy,epsfig,color,graphicx,times}
\usepackage[ansinew]{inputenc}
\usepackage[english]{babel}
\usepackage{color}
\usepackage{braket}
\usepackage{tabularx}
\usepackage{array}

\begin{document}

\title{Quantum reference frames and particle mixing}

\author{Antonio Capolupo$^{1}$ and Aniello Quaranta$^{2}$}

\affil{$^1$Dipartimento di Fisica ``E.R. Caianiello'' and INFN gruppo collegato di Salerno, Universit\`{a} degli Studi di Salerno, I-84081
Fisciano (SA), Italy}
\affil{$^2$School of Science and Technology, University of Camerino,
Via Madonna delle Carceri, Camerino, 62032, Italy}

\email{capolupo@sa.infn.it, aniello.quaranta@unicam.it}

\begin{abstract} We discuss the necessity and the emergence of quantum reference frames when attempting to define a rest frame for mixed particles. We analyze the corresponding concept of frame dependent entanglement and how it could affect measurements on mixed mesons and neutrinos \cite{QR1}.
\end{abstract}

\section{Introduction}
 
 The principle of relativity is one of the cornerstones of Physics. The evolution of the concept of reference frame, along with the transformations connecting different frames, has led some of the most important developments within the foundations of Physics. The revolution brought forward by Special Relativity, for instance, relies on a fundamental upgrade in the notion of reference frames, by the addition of clocks as a necessary element, and on the replacement of Galileian transfromations with Lorentzian transformations as appropriate for the new 4-dimensional scenario.
 By definition a reference frame is the set of references that we employ to describe physics, namely a spacetime event to be used as origin and a coordinate system. To this rather abstract notion corresponds the actual physical practice to pinpoint a \emph{physical system} to serve as reference, along with (generalized) clocks and rods, with respect to which the physical phenomena are described.
 
 By now there is a large consensus regarding the fundamental nature of physical systems as quantum mechanical. It is therefore natural to demand for an upgrade in our notion of reference frame to capture the quantumness of the object that serves as reference for the description of Physics. \emph{Quantum ReferenceFrames} (QRFs) have first appeared in the context of quantum information  \cite{QI1,QI2,QI3,QI4,QI5,QIR,QI6,QI7,QI8,QI9,QI10,QI11,QI12}, and their significance in Quantum Gravity was soon recognized \cite{QG1,QG2,QG3,QG4}, as pertaining to the issue of relational observables. Following the relational paradigm \cite{QIR,QIR2}, a foundational approach to quantum reference frames was developed \cite{FQR1,FQR2,FQR3,FQR4,FQR5,FQR6,FQR7,FQR8,FQR9,FQR10}.

 In this paper, following the analysis of \cite{QR1}, we argue that QRFs are necessary to provide a full description of mixed particle systems (neutrinos, neutral mesons, etc.). We show, in particular, that the rest frame for a mixed particle is classically ill-defined, making sense only as a QRF. We further discuss how one of the most striking peculiarities of QRFs, namely the relativity of entanglement, can affect reactions involving mixed particles.
 
 \section{Flavor neutrino states}
 
 After Wigner \cite{Wigner} we do recognize \emph{fundamental particles} to belong to irreducible projective representations of the Poincar\`{e} group. The one-particle states of (fundamental) quantum field theories belong to any of the irreducible representations $\ket{m,s}$, labeled by mass $m$, the eigenvalue of the Casimir operator $P^\mu P_{\mu}$ and the spin $s$, that for massive $m>0$ particles is related to the eigenvalues of the Pauli-Lubanski pseudovector as $W^\mu W_\mu \ket{m,s} = -m^2 s(s+1) \ket{m,s}$. For most applications it is convenient to employ the momentum basis  $\ket{p^{\mu}, \sigma }_{m,s}$, provided by the simultaneous eigenvectors of the momentum (translation) operators $P^{\mu} \ket{p^{\mu}, \sigma }_{m,s} = p^{\mu} \ket{p^{\mu}, \sigma }_{m,s}$. The Poincar\`{e} group acts unitarily on these states \cite{Wigner,Weinberg}: $U (\Lambda) \ket{p^{\mu}, \sigma }_{m,s} = \sum_{\sigma'} D^{(s)}_{\sigma' \sigma} (W(\Lambda, p)) \ket{\Lambda^{\mu}_{\nu} p^{\nu}, \sigma' }_{m,s} $  and $U(a) \ket{p^{\mu}, \sigma }_{m,s} = e^{-i p^{\mu} a_{\mu}} \ket{p^{\mu}, \sigma }_{m,s}$. Here $\Lambda$ stands for the homogeneous Lorentz transformations, $a^{\mu}$ is the translation $4$-vector, the rotation coefficients $D^{s}_{\sigma' \sigma} (R)$ implement the $2s + 1$-dimensional unitary irreducible representation of the rotation group (we are assuming $m > 0$), and $W(\Lambda,p)$ is the Wigner rotation. 
 Obviously no two sectors $\ket{m,s}$ and $\ket{m',s'}$ with $m \neq m'$ or $s \neq s'$ can ever be connected by a Poincar\`{e} transformation.
 
Flavor neutrinos \cite{Neut1,Neut2,Neut3,Neut4,Neut5,Neut6,Neut7,Neut8,Neut9,Neut10,Neut11,Neut12,Neut13,Neut14,Neut15,Neut16} represent the only known exception to this paradigm. They are considered as elementary particles, but at the same time they do not, at least strictly speaking, belong to an irreducible representation $\ket{m,s}$. For notational simplicity, in the following, we shall omit the spin label, as it is irrelevant for the upcoming discussion. 
The state of an electron neutrino may be written, after Pontecorvo, as the superposition 
\begin{equation}\label{NeutrinoState}
 \ket{\pmb{p}, e} = \cos \theta \ket{p^{\mu}_{1}}_{m_1} + \sin \theta \ket{p_{2}^{\mu}}_{m_2}
\end{equation}
where we have limited to two flavors. The momenta are specified as $p_{j}^{\mu} \equiv \left( \pmb{p}, \omega_j \right)$ with the on-shell energies $\omega_j = \sqrt{\pmb{p}^2 + m_j^2}$. Clearly the states of Eq. \eqref{NeutrinoState} are eigenstates of the $3$-momentum $\pmb{P} \ket{\pmb{p},e} = \pmb{p} \ket{\pmb{p},e}$, but not eigenstates of the energy $P^0$. A similar expression holds for the muon neutrino state. As a consequence the state of Eq. \eqref{NeutrinoState} transforms non-trivially under any transformation involving the temporal direction. For instance a translation by $a^{\mu} \equiv (a,0,0,0)$ gives 
\begin{equation}
 U(a) \ket{\pmb{p}, e} = \cos \theta e^{-i\omega_1 a} \ket{p^{\mu}_{1}}_{m_1} + \sin \theta e^{-i\omega_2 a} \ket{p_{2}^{\mu}}_{m_2} \ .
\end{equation}
The misalignment in the phase factors is what sources the flavor oscillations in the free propagation of neutrinos. 

From the above equations it is clear that flavor neutrino states always involve two (three for three flavor mixing) distinct irreducible representations with different masses $m_1, m_2$. Besides prompting the oscillation phenomenon, and rendering neutrinos unique among elementary particles, this simple fact has a significant impact on the interaction processes involving neutrinos \cite{QR1}.

\subsection{The rest frame of flavor neutrinos}

Particles with definite momentum $\ket{p^{\mu}}_{m}$ admit a natural notion of rest frame. Up to spatial rotations, it is the unique frame in which the particle's momentum takes on the rest form $\ket{p^{\mu}_{(0)}}_{m} \equiv \ket{(\pmb{0},m) }_{m}$. Given an arbitrary frame in which the particle has momentum $p^{\mu}$, there is a unique Lorentz transformation $\Lambda$ (again, up to spatial rotations) that takes the particle to its rest frame:
\begin{equation}
 p^{\mu}_{{0}} = \Lambda^{\mu}_{\nu} p^{\nu} \ .
\end{equation}
If we assume, without loss of generality, that the spatial momentum is aligned with the $z$ axis, the transformation is simply the Lorentz boost
\begin{equation}\label{ClassicalTransformation}
 \Lambda^{\mu}_{\nu} \equiv \begin{pmatrix} \frac{\sqrt{p^2 + m^2}}{m} & 0 & 0 & \frac{-p}{m} \\ 0 & 1 & 0 & 0 \\ 0 & 0 & 1 & 0 \\ \frac{-p}{m} & 0 & 0 & \frac{\sqrt{p^2 + m^2}}{m} \end{pmatrix} \ .
\end{equation}
Notice that the boost parameter $\psi = \cosh^{-1} \left( \frac{\sqrt{p^2 + m^2}}{m} \right)$ is a $c$-number. 
Can we define a notion of rest frame for a particle, as the flavor neutrino of Eq. \eqref{NeutrinoState} that does not possess a definite momentum? To do so, we need to determine a frame in which the particle is at rest, that is, in which its $3$-momentum vanishes:
\begin{equation}\label{NeutrinoRestState}
 \ket{\pmb{0}, e} = \cos \theta \ket{\left(\pmb{0}, m_1 \right)}_{m_1} + \sin \theta \ket{\left(\pmb{0}, m_2 \right)}_{m_2}
\end{equation}
It is easy to check that there does not exist a classical Lorentz boost of the form \eqref{ClassicalTransformation} able to simultaneously annihilate the $3$-momentum of the $m_1$ and $m_2$ components of the neutrino state. This is simply due to the mismatch in the boost parameters needed to bring $p_1^{\mu}$ and $p_2^{\mu}$ to their rest form. Albeit no classical transformation can lead from the state of Eq. \eqref{NeutrinoState} to the rest state of Eq. \eqref{NeutrinoRestState}, a quantum transformation can. To see this, promote the boost parameter to an operator 
\begin{equation}
\hat{\psi} = \cosh^{-1} \left( \hat{H}\left( \hat{P}^{\mu} \hat{P}_{\mu} \right)^{-\frac{1}{2}} \right) \ ,
\end{equation}
 with $\hat{H} = \hat{P}^0$, and the hats have been introduced to underline the operatorial nature. In analogy with the classical transformation \eqref{ClassicalTransformation} introduce the quantum Lorentz boost
\begin{equation}\label{QuantumTransform}
  \hat{\Lambda}^{\mu}_{\nu} \equiv \begin{pmatrix}  \hat{H}\left( \hat{P}^{\mu} \hat{P}_{\mu} \right)^{-\frac{1}{2}} & 0 & 0 & -\hat{P}_z\left( \hat{P}^{\mu} \hat{P}_{\mu} \right)^{-\frac{1}{2}} \\ 0 & 1 & 0 & 0 \\ 0 & 0 & 1 & 0 \\ -\hat{P}_z\left( \hat{P}^{\mu} \hat{P}_{\mu} \right)^{-\frac{1}{2}} & 0 & 0 &  \hat{H}\left( \hat{P}^{\mu} \hat{P}_{\mu} \right)^{-\frac{1}{2}} \end{pmatrix}  \
\end{equation}
The corresponding unitary representation $U(\hat{\Lambda})$ transforms the neutrino state of Eq. \eqref{NeutrinoState} (here we have assumed the spatial momentum along $z$) into its rest form (Eq. \eqref{NeutrinoRestState}). Obviously, upon acting on a state with definite $4$-momentum, the quantum boost reduces to the classical Lorentz boost.
The quantum boost $U(\hat{\Lambda})$ is the same kind of transformation that occurs in moving from a QRF to another \cite{FQR2,FQR4}. The need for quantum transformations is not limited to the rest frame of the flavor neutrinos. Suppose, more generally, that in a given frame the neutrino is characterized by the $3$-momentum $\pmb{p}$. The transformation needed to move to a frame in which the neutrino has $3$-momentum $\pmb{p}' \neq \pmb{p}$ is, assuming both $\pmb{p}'$ and $\pmb{p}$ along the third axis, the quantum Lorentz boost with parameter 
\begin{equation}
 \hat {\psi} = \sinh^{-1} \left( \left(\hat{P}^{\mu} \hat{P}_{\mu}\right)^{-1} \left(p' \hat{H} - \sqrt{p^{'2} + \hat{P}^{\mu} \hat{P}_{\mu}} \hat{P}_z \right) \right) \ .
\end{equation}
It is clear that to make any sense of a neutrino rest frame, and, more generally, to correctly associate reference frames to mixed particles, the concept of QRFs and QRF transformations are unavoidable. 

An important remark is now in order. A quantum Lorentz transformation is needed to the define the rest frame for any particle that does not possess a definite $4$-momentum (see for instance \cite{FQR2}). This is the case also for generic wave packets $\ket{\phi} = \int d^4 p \ \theta (p^0) \ \delta (p^{\mu} p_{\mu} - m^2) \phi (p) \ket{p^{\mu}}_{m}$, even for particles that are not mixed (just a single definite value of mass).
However, neutrino flavor states are different from wave packets in that the QRFs transformations are \emph{fundamentally} needed to define their rest frame. Firstly, no matter how sharply defined one may get the neutrino $3$-momentum to be, the flavor state will always involve at least a superposition of two states, belonging to two distinct irreducible representations of masses $m_1$ and $m_2$. A flavor state can \emph{never} be a simultaneous eigenstate of all the translation operators $P^\mu$. Secondly, for (free) elementary particles there is no \emph{fundamental} reason for which they cannot be in a sharp $4$-momentum eigenstate. On the opposite, flavor neutrinos do always necessarily involve a momentum superposition and carry an essential momentum uncertainty.

\section{Relativity of entanglement for mixed particles}

The conclusions drawn in the previous section hold for any kind of mixed particle, be they fundamental (neutrinos) or composite (neutral mesons \cite{BM1,BM2,BM3,BM4,BM5,BM6,BM7,BM8,BM9,BM10,BM11,BM12} and neutron-mirror neutron mixing \cite{NMN1,NMN2,NMN3}). The hallmark of QRFs is the \emph{relativity of entanglement}: quantum entanglement depends on the specific QRF. To illustrate this concept in its simplest realization, consider three particles $A,B,C$ arranged on a line. Following \cite{FQR2,FQR5}, denote the state of particle $J$ with respect to the QRF of particle $K$ (i.e. the QRF in which $K$ is permanently at the origin, $x_K = 0$) as $\ket{x_J}^{(K)}_J$. Suppose that the state of $B$ and $C$ in the QRF of $A$ has the form
\begin{equation}
 \psi^{(A)}_{BC} = \ket{x_B}_B^{(A)} \left(\ket{x_C^1}_C^{(A)} + \ket{x_C^2}_C^{(A)} \right) \ .
\end{equation}
Here $C$ is in a superposition of different locations $x^1_C$ and $x^2_C$. Notice that the state is factorizable (zero entanglement). The state of $A$ with respect to this frame is obviously $\ket{0}_A^{(A)}$.
Let us now move to the QRF of $C$. In order to do so we need to implement a quantum translation by $-\hat{x}_C$ and the resulting state of $A,B$ with respect to $C$ reads
\begin{eqnarray}\label{EntangledState}
 \nonumber && \psi^{(C)}_{AB} = \left(T(-x^1_C) + T(-x^2_C) \right) \left(\ket{x_A = 0}_A \ket{x_B}_B \right) \\
 && = \ket{-x^1_C}_A^{(C)} \ket{x_B -x^1_C}_B^{(C)}  +  \ket{-x^2_C}_A^{(C)} \ket{x_B -x^2_C}_B^{(C)} \ ,
\end{eqnarray}
where $T(a)$ denotes an ordinary translation by $a$. Due to the original superposition of $C$ positions, the quantum translation has turned a factorizable state into an entangled state, namely Eq. \eqref{EntangledState}.
Let us now see the relativity of entanglement in action for mixed particles. We revert to the $4$-momentum basis and consider three particles $A,B$ and a flavor neutrino $\nu$. Let $B$ be in a sharp momentum eigenstate and $\nu$ be in the generic electron neutrino state \eqref{NeutrinoState}, as seen by $A$:
\begin{equation}\label{Astate}
 \psi_{B\nu}^{(A)} = \ket{p_B^{\mu}}_B^{(A)} \left( \cos \theta \ket{p^{\mu}_{1}}_{m_1, \nu}^{(A)} + \sin \theta \ket{p_{2}^{\mu}}^{(A)}_{m_2, \nu} \right) \ .
\end{equation}
Analogously to the case of translations, the omitted rest state of $A$ with respect to $A$ is $\ket{p^{\mu}_{A,0}}_A^{(A)} = \ket{(\pmb{0},m_A)}_A^{(A)}$.
What is the state of $A,B$ with respect to $\nu$? To answer this question we perform the Quantum Lorentz boost of Eq. \eqref{QuantumTransform} to get to the rest frame of $\nu$, so to get
\begin{eqnarray}\label{NeutrinoEntangledState}
 \nonumber  \psi_{AB}^{(\nu)} &=& U(\hat{\Lambda})\left( \ket{p^{\mu}_{A,0}}_A \ket{p^{\mu}_B}_B\right)= \left(\cos \theta  \ U(\Lambda_1) + \sin \theta \ U(\Lambda_2) \right) \left( \ket{p^{\mu}_{A,0}}_A \ket{p^{\mu}_B}_B\right)  \\ 
 &=&  \cos \theta \ket{\Lambda^{\mu}_{1,\nu}p_{A,0}^{\nu}}_A^{(\nu)} \ket{\Lambda^{\mu}_{1,\nu}p_{B}^{\nu}}_B^{(\nu)}
+\sin \theta \ket{\Lambda^{\mu}_{2,\nu}p_{A,0}^{\nu}}_A^{(\nu)} \ket{\Lambda^{\mu}_{2,\nu}p_{B}^{\nu}}_B^{(\nu)} \ .
\end{eqnarray}
Clearly the operators $U(\Lambda_j)$ act on the states of $A$ and $B$, and $\Lambda_j$ denote the (classical) Lorentz boosts of Eq. \eqref{ClassicalTransformation} corresponding to masses $m_1$ and $m_2$. The state \eqref{NeutrinoEntangledState} is manifestly entangled.
\subsection{Entanglement in neutral meson decays}
To show how the frame related entanglement affects actual physical processes, let us consider the decay of a charged $D$ meson into three particles \cite{PDG} $D^{+} \rightarrow \bar{K}^0 + e^+ + \nu_e $. The strangeness eigenstate (neglecting CP violation for simplicity) $\ket{\bar{K}^0} = \frac{1}{\sqrt{2}}\left(\ket{K_L} - \ket{K_S} \right)$ is a linear combination of the mass eigenstates $\ket{K_L}, \ket{K_S}$, the latter corresponding to definite masses $m_L, m_S$. For the quantum mechanical analysis conducted here, where spin is neglected, $\ket{\bar{K}^0}$ is perfectly analogous to the flavor neutrino state of Eq. \eqref{NeutrinoState}. If the Kaon has definite $3$-momentum $\pmb{k}$ in the laboratory frame, we may write the state of the decay products as
\begin{eqnarray}
 \nonumber &&\ket{DP}^{LAB} =  \frac{1}{\sqrt{2}}\left(\ket{(\pmb{k}, \omega_L)}_{m_L, K}^{LAB} - \ket{(\pmb{k}, \omega_S)}_{m_S, K}^{LAB} \right)\ket{p^{\mu}_{e^+}}_{e^+}^{LAB} \ket{p^{\mu}_{\nu_e}}_{\nu_e}^{LAB}  \ , 
\end{eqnarray}
where $\omega_{L,S} = \sqrt{\pmb{k}^2 + m_{L,S}^2}$. Now consider the same decay, characterized by the same kinematical Mandelstam invariants $s,u,t$, but suppose that the Kaon is nearly at rest in this frame, so that it may be possible to identify the laboratory frame with the QRF of the Kaon. In this case the state of the decay products reads
\begin{eqnarray}\label{EntangledKaon}
\ket{DP}^{(K)} =    \frac{\Big(\ket{\Lambda^{\mu}_{L,\nu}p^{\nu}_{e^+}}_{e^+}^{(K)} \ket{\Lambda^{\mu}_{L,\nu}p^{\nu}_{\nu_e}}_{\nu_e}^{(K)}\ket{(\pmb{0},m_L)}_{m_L, K}^{(K)} - \ket{\Lambda^{\mu}_{S,\nu}p^{\nu}_{e^+}}_{e^+}^{(K)} \ket{\Lambda^{\mu}_{S,\nu}p^{\nu}_{\nu_e}}_{\nu_e}^{(K)}\ket{(\pmb{0},m_s)}_{m_s, K}^{(K)} \Big)}{\sqrt{2}} \ .
\end{eqnarray}
The above state displays entanglement among the leptons, whereas they were originally in a product state in the laboratory frame. We stress that this form of entanglement is exclusively related to QRF transformations.

\subsection{Possible observable signatures}
Although the entanglement itself is frame-dependent, behind the scenes all the transformations involved are unitary, as extensively discussed for instance in \cite{FQR7}. Entanglement-related observables are simply mapped (unitarily) into distinct observables by QRF transformations. To illustrate this point, consider the states of Eqs. \eqref{Astate} and \eqref{NeutrinoEntangledState} and denote by $\mathcal{H}^{(K)}_{J}$ the Hilbert space of particle $J$ in the QRF of particle $K$. The quantum boost transforming from the $A$ frame to the $\nu$ frame is the unitary map
\begin{equation}
 U(\hat{\Lambda}) : \mathcal{H}^{(A)}_B \otimes \mathcal{H}^{(A)}_{\nu} \rightarrow \mathcal{H}_A^{(\nu)} \otimes \mathcal{H}_B^{(\nu)}
\end{equation}
that reads the neutrino momentum as input (via $\hat{\Lambda}$) and performs the corresponding boost on $A$ and $B$. We will now drop the argument of $U$ for notational convenience. If $O^{(\nu)}_{AB} : \mathcal{H}_A^{(\nu)} \otimes \mathcal{H}_B^{(\nu)} \rightarrow \mathcal{H}_A^{(\nu)} \otimes \mathcal{H}_B^{(\nu)}$ is an observable in the rest frame of the flavor neutrino $\nu$, then the inverse QRF transformation $U^{\dagger}$ maps it unitarily to an observable in the $A$ frame as \cite{FQR2}:
\begin{equation}
 O^{(A)}_{B\nu} = U^{\dagger}O^{(\nu)}_{AB}U \ .
\end{equation}
Then, by unitarity
\begin{equation}\label{Unitarity}
\bra{\psi^{(A)}_{B\nu}} O^{(A)}_{B\nu} \ket{\psi^{(A)}_{B\nu}} = \bra{\psi^{(\nu)}_{AB}} O^{(\nu)}_{AB} \ket{\psi^{(\nu)}_{AB}} \ ,
\end{equation}
or, equivalently, $\mathrm{Tr}_{AB}(\rho^{(\nu)}_{AB}O^{(\nu)}_{AB}) = \mathrm{Tr}_{B\nu}(\rho^{(A)}_{B\nu}O^{(A)}_{B\nu})$, in terms of the density matrices $\rho^{(\nu)}_{AB}$ and $\rho^{(A)}_{B\nu}$.
This is peculiar to QRFs: the two sides of Eq. \eqref{Unitarity} have the same numerical value, but a completely different interpretation within the two frames. The QRFs, in particular, do even disagree on the system with respect to which the observables are referred to ($B\nu$ or $AB$). At the same time, unitarity, in the shape of Eq. \eqref{Unitarity}, implies that we observe as entanglement in a frame is just transformed into a distinct quantity (not necessarily related to entanglement) by the QRF transformation.
To make this point even clearer, consider the entangled state of Eq. \eqref{NeutrinoEntangledState} and define an entanglement witness, that is, an observable which vanishes in absence of entanglement and signals its presence. Denote, for notational simplicity, $\bar{p}^{\mu}_{A,1}=\Lambda^{\mu}_{1,\nu}p^{\nu}_{A,0}, \bar{p}^{\mu}_{A,2}=\Lambda^{\mu}_{2,\nu}p^{\nu}_{A,0}, \bar{p}^{\mu}_{B,1}=\Lambda^{\mu}_{1,\nu}p^{\nu}_{B}, \bar{p}^{\mu}_{B,2}=\Lambda^{\mu}_{2,\nu}p^{\nu}_{B}$, and rewrite the state \eqref{NeutrinoEntangledState} as
\begin{equation}
 \psi^{(\nu)}_{AB} = \cos \theta \ket{\bar{p}^{\mu}_{A,1}}_A^{(\nu)} \ket{\bar{p}^{\mu}_{B,1}}_B^{(\nu)} + \sin \theta \ket{\bar{p}^{\mu}_{A,2}}_A^{(\nu)} \ket{\bar{p}^{\mu}_{B,2}}_B^{(\nu)} \ .
\end{equation}
Let $L_{AB}^{(\nu)}$ be the tensor product operator
\begin{equation}
 L_{AB}^{(\nu)} = L_A \otimes L_B : \mathcal{H}_A^{(\nu)} \otimes \mathcal{H}_B^{(\nu)} \rightarrow \mathcal{H}_A^{(\nu)} \otimes \mathcal{H}_B^{(\nu)} 
\end{equation}
where each of the operators $L_J$ has the only nonvanishing matrix elements
\begin{equation}
\bra{\bar{p}^{\mu}_{J,2}} L_{J} \ket{\bar{p}^{\mu}_{J,1}} = \epsilon_{J} 
\end{equation}
for $J = A,B$ and some complex constants $\epsilon_{J}$. The operator defined above may be related to some interaction Hamiltonian driving momentum transitions in $A$ and $B$. It is straightforward to check that $L_{AB}^{(\nu)}$ is an entanglement witness. On product states one does indeed find

\begin{equation}
 \bra{\bar{p}^{\mu}_{A,j}} \bra{\bar{p}^{\mu}_{B,k}} L^{(\nu)}_{AB} \ket{\bar{p}^{\mu}_{B,k}} \ket{\bar{p}^{\mu}_{A,j}} = 0
\end{equation}
for all $j,k=1,2$. On the other hand, a simple computation shows that
\begin{equation}\label{Witness}
\bra{\psi_{AB}^{(\nu)}} L^{(\nu)}_{AB} \ket{\psi_{AB}^{(\nu)}} = 2 \cos\theta \sin \theta \ \mathrm{Re} \left(\epsilon_A \epsilon_B \right) \ .
\end{equation}
Comparing the Eq. \eqref{Witness} to the linear entropy $E_L = 1- \mathrm{Tr}(\rho^2_R)$, where $\rho_R$ is the reduced density matrix with respect to $A$ or $B$, we find that
\begin{equation}
 E_L = \frac{\bra{\psi_{AB}^{(\nu)}} L^{(\nu)}_{AB} \ket{\psi_{AB}^{(\nu)}}^2}{2 \mathrm{Re}^2 \left( \epsilon_A \epsilon_B\right)} \ .
\end{equation}
Therefore the linear entropy vanishes whenever the witness does, as required. In the neutrino frame, the observable \cite{Witness} is directly related to the entanglement in the $AB$ state. Transforming back to the $A$ frame, in accordance with the general prescription of Eq. \eqref{Unitarity}, the witness \eqref{Witness}is turned into the expectation value
\begin{equation}
 \bra{\psi^{(A)}_{B\nu}} O^{(A)}_{B\nu} \ket{\psi^{(A)}_{B\nu}} = 2 \cos \theta \sin \theta \mathrm{Re} \left(\epsilon_A \epsilon_B \right) \ .
\end{equation}
Clearly $O^{(A)}_{B\nu}$ is not related to any entanglement, which is absent for the product state of eq. \eqref{Astate}. Nonetheless unitarity provides, in this way, a means to detect the frame-related entanglement, even within a frame where entanglement is absent, by simply measuring the appropriate observable.  

\section{Conclusions}
QRFs have emerged, in the latest years, for their primary importance in relational descriptions of Quantum Gravity, where even the notion of a classical reference frame may fail to be appropriate. In this work, following the reference \cite{QR1} we have shown that QRFs may be fundamental also in particle physics, specifically pertaining to particle mixing. We have argued that the notion of a rest frame for a mixed particle necessarily falls within the context of QRFs. We have then displayed how relativity of entanglement, a striking signature of QRFs, may have an observable impact on the dynamics of mixed particle systems (neutral mesons, neutrinos, etc.)

\section*{Acknowledgements}
Partial financial support from MIUR and INFN is acknowledged.
A.C. also acknowledges the COST Action CA1511 Cosmology
and Astrophysics Network for Theoretical Advances and Training Actions (CANTATA).

\end{document}